\documentclass[twocolumn,trackchanges]{aastex631}
\usepackage{comment}

\newcommand\nustar{\textit{NuSTAR}}
\newcommand\hic{\textit{Hi-C 2.1}}
\newcommand\foxsi{\textit{FOXSI}}
\newcommand\rhessi{\textit{RHESSI}}

\def\apjl{ApJL}                  			% Astrophysical Journal, Letters
\def\apjs{ApJS}       		 		% Astrophysical Journal, Supplement
\def\aap{A\&A}                   		% Astronomy and Astrophysics
\received{December 8, 2023}
\submitjournal{ApJ}

\shortauthors{Duncan et al.}
\graphicspath{{./}{figures/}}
\begin{document}

%\title{Hard X-ray Diagnostics of Thermal Plasma in an Alternately Quiet \& Microflaring Active Region}
\title{Thermal Evolution of an Active Region through Quiet and Flaring Phases as Observed by \nustar\, XRT, and AIA}

\author[0000-0002-6872-4406]{Jessie Duncan}
\affiliation{NASA Postdoctoral Program}
\affiliation{NASA Goddard Space Flight Center, Greenbelt, MD, USA}

\author[0000-0003-2395-9524]{Reed B. Masek}
\affiliation{University of Minnesota, Minneapolis, MN, USA}

\author[0000-0001-6874-2594]{Albert Y. Shih}
\affiliation{NASA Goddard Space Flight Center, Greenbelt, MD, USA}

\author[0000-0001-7092-2703]{Lindsay Glesener}
\affiliation{University of Minnesota, Minneapolis, MN, USA}

\author[0000-0001-9642-6089]{Will Barnes}
\affiliation{NASA Goddard Space Flight Center, Greenbelt, MD, USA}
\affiliation{American University, Washington, DC, USA}

\author[0000-0002-6903-6832]{Katharine K. Reeves}
\affiliation{Harvard-Smithsonian Center for Astrophysics, Cambridge, MA, USA}

\author[0000-0001-8941-2017]{Yixian Zhang}
\affiliation{University of Minnesota, Minneapolis, MN, USA}

\author[0000-0003-1193-8603]{Iain G. Hannah}
\affiliation{University of Glasgow, Glasgow, UK}

\author[0000-0002-1984-2932]{Brian W. Grefenstette}
\affiliation{California Institute of Technology, Pasadena, CA, USA}



\begin{abstract}

Solar active regions contain a broad range of temperatures, with the thermal plasma distribution often observed to peak in the few millions of kelvin. Differential emission measure (DEM) analysis can allow instruments with diverse temperature responses to be used in concert to estimate this distribution. \nustar\ HXR observations are uniquely sensitive to the highest-temperature components of the corona, and thus extremely powerful for examining signatures of reconnection-driven heating. Here, we use \nustar\ diagnostics in combination with EUV and SXR observations (from \textit{SDO}/AIA and \textit{Hinode}/XRT) to construct DEMs over 170 distinct time intervals during a five-hour observation of an alternately flaring and quiet active region (NOAA designation AR 12712). This represents the first HXR study to examine the time evolution of the distribution of thermal plasma in an active region. During microflares, we find that the initial microflare-associated plasma heating is dominantly heating of material that is already relatively hot, followed later on by broader heating of initially-cooler material. During quiescent times, we show that the amount of extremely hot ($>$10 MK) material in this region is significantly ($\sim$3 orders of magnitude) less than that found in the quiescent active region observed in HXRs by \foxsi-2 \citep{2017NatAs...1..771I}. This result implies there can be radically different high-temperature thermal distributions in different active regions, and strongly motivates future HXR DEM studies covering a large number of these regions.

\end{abstract}

\section{Introduction} \label{sec:intro}

Plasma in the solar corona is persistently observed to be much hotter than the $\sim$6000 K photosphere, with active regions (ARs) typically found to be at temperatures of 2–3 MK or greater. The source of energy leading to this elevated temperature must be the solar magnetic field. It is so far unclear which mechanism is dominant in converting magnetic energy into consistent coronal heating.

During solar flares, energy is released through magnetic reconnection, as coronal magnetic fields abruptly relax to a lower-potential state. Significant heating of AR plasma occurs in flares. However, observed solar flares occur insufficiently often to explain the persistently elevated temperature of the corona as a whole \citep[e.g.,][]{1995PASJ...47..251S}.

Reconnection still may lead to coronal heating— instead, via a large ensemble of ``nanoflares", tiny events too faint to individually observe \citep{1988ApJ...330..474P}. It is unknown exactly how released magnetic energy would be converted to plasma heating in these events; one method for investigating this mechanism is the continued study of energy release and plasma heating in microflares, the smallest flares we \textit{can} individually observe \citep[e.g.,][]{2008ApJ...677.1385C, 2008ApJ...677..704H, 2020ApJ...891...78A, 2021MNRAS.507.3936C}. One consistent property is that the hottest flare-heated plasma is present early in the evolution of the event, more closely linked to the initial energy release. 

Nanoflare (or low-frequency coronal) heating can be contrasted with theories of heating via magnetohydrodynamic (MHD) wave dissipation, which involve a continuous input of energy into the corona \citep[high-frequency heating;][]{2020SSRv..216..140V}. An unambiguous observational signature of low-frequency heating is the presence of faint, extremely hot plasma ($>$7 MK) occurring in the absence of any observable impulsive event \citep{1994ApJ...422..381C, cargill_nanoflare_2004}. 

To characterize plasma heating at quiet \textit{or} flaring times, it is desirable to determine the distribution of thermal plasma present as a function of time— particularly the highest-temperature material. Differential Emission Measure (DEM) analysis allows estimation of the emission measure distribution of a source as a function of temperature, starting from observation(s) of a source and the temperature response(s) of the observing instrument(s).  

There have been a wealth of quiescent AR DEM studies \citep[summarized, for example, in][]{2016ApJ...829...31B} utilizing extreme ultraviolet (EUV) and soft X-ray (SXR) diagnostics from \textit{Hinode}'s EUV Imaging Spectrometer (EIS) and X-Ray Telescope (XRT), as well as the Atmospheric Imaging Assembly (AIA) aboard the \textit{Solar Dynamics Observatory} (\textit{SDO}) \citep{2007SoPh..243...19C, XRT_instrument, AIA_instrument}. However, the combination of these instruments is insufficiently sensitive to faint components above 5 MK \citep{Winebarger_2012}. Additionally, especially short-lived hot plasma may not produce signatures in the hot EUV spectral lines observed by EIS and AIA, due to a lack of ionization equilibrium \citep[e.g.,][]{2003A&A...401..699B,2011ApJS..194...26B}. 

A powerful alternate diagnostic for constraining high temperature material is the thermal bremsstrahlung continuum as observed in the hard X-ray (HXR) range \citep[e.g.,][]{Ishikawa_2019}. Only the high-energy tail of the electron distribution generates bremsstrahlung at HXR energies, biasing this diagnostic toward the hottest material. Additionally, continuum observations are much less sensitive to non-equilibrium ionization effects.

The \textit{Reuven Ramaty High Energy Solar Spectroscopic Imager} (\rhessi) was a flare-optimized HXR indirect imaging spectrometer. \rhessi's indirect imaging method gave it limited sensitivity, due to a high non-solar background. However, \rhessi\ was still useful for constraining or identifying high-temperature material. Considering quiescent times, \cite{2009ApJ...704..863S} used \rhessi\ to significantly constrain the hot plasma possible in a quiescent AR, when compared to results with EUV/SXR instruments alone. Additionally, \rhessi\ estimates of the plasma emission measure have identified the presence of a component above 5 MK at non-flaring times \citep{McTiernan_2009, 2009ApJ...704L..58R}. Finally, \rhessi\ can be incorporated in DEM analyses: \cite{Ishikawa_2019} used \rhessi, XRT, and AIA to derive the DEM of a quiet active region, unambiguously identifying a faint $>$5 MK component. Unlike the majority of quiescent ARs, this region was bright enough in HXRs to be imaged by \rhessi\ (albeit via a long integration) — likely indicating it was also hotter than the typical AR. That only particularly bright regions can be analyzed this way by \rhessi\ represents an observational bias. 

Considering small transients, a large two-part \rhessi\ study examined flares from GOES A to C-class, with the second paper utilizing \rhessi\ spectroscopy to extract isothermal approximations of the plasma heated in each \citep{2008ApJ...677.1385C, 2008ApJ...677..704H}. Additionally, \cite{2014ApJ...789..116I} conducted a joint AIA-\rhessi\ analysis of ten B-class microflares, in which forward-fitting of AIA data and \rhessi\ spectroscopy were combined to estimate parameters describing DEMs of an assumed functional form. 

%==================================================================
%==================================================================
 \begin{figure*}[!ht] 
\centering 
\includegraphics[width=\textwidth]{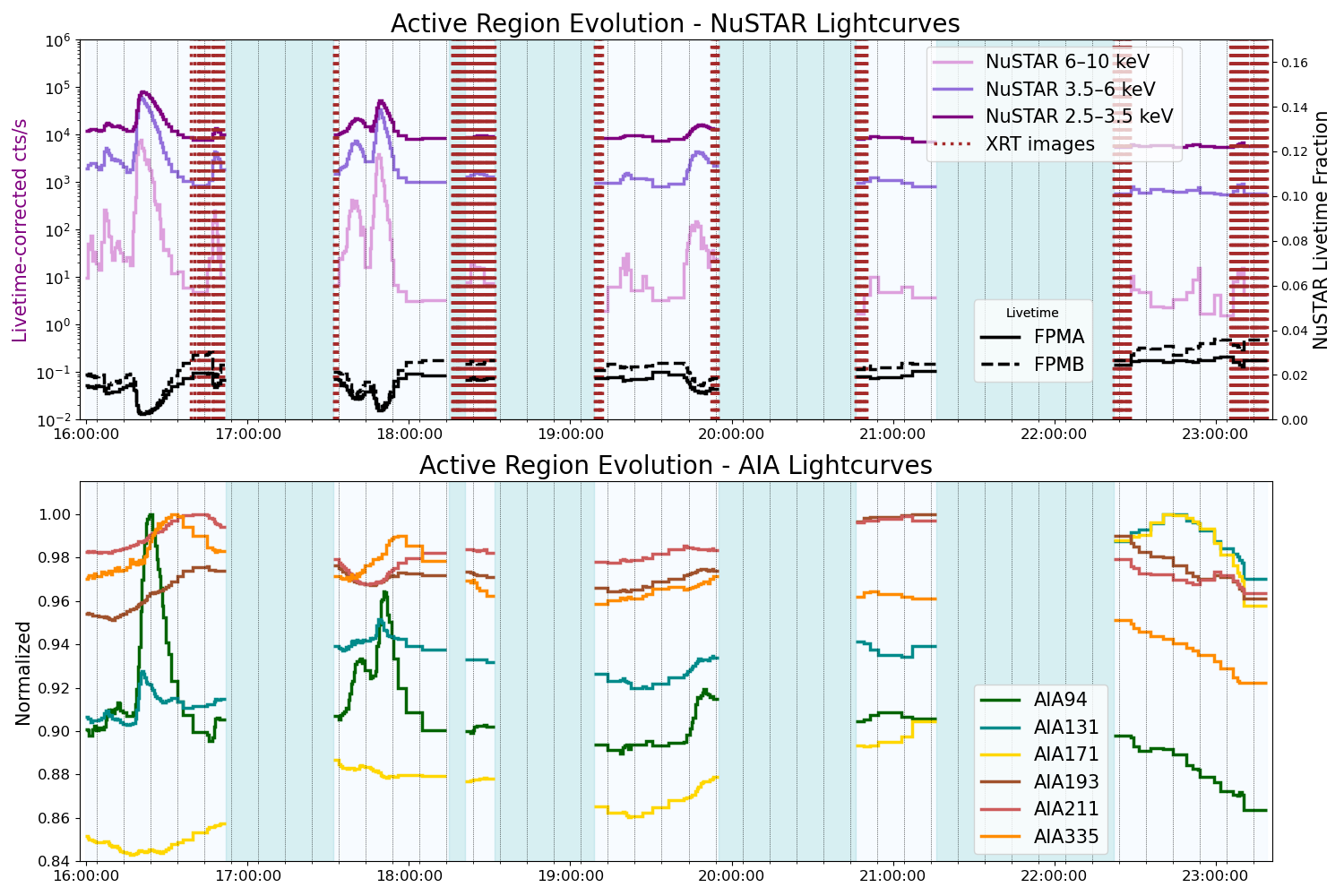} 
\caption{Time evolution of AR 12712 during the \nustar\ observation intervals. Time binning of lightcurves is according to the DEM time intervals (described in Section \ref{timeselect}). \textbf{Upper:} The top 3 curves show \nustar\ livetime-corrected lightcurves in the three energy ranges used in DEM analysis, scale on left axis. \nustar\ livetime-corrected counts have been summed between its two telescopes (FPMA, FPMB, see Section \ref{nuprep}). XRT observation times during the \nustar\ observations are marked. The lower two black curves show the livetime of each \nustar\ telescope during the same intervals, scale on right axis. \textbf{Lower:} AIA DEM inputs are shown for each DEM time interval; see Section \ref{aiaprep} for details on data preparation. In this plot, the values in each AIA channel are normalized to their maximum over the entire duration of the observation.} 
\label{fig:lcoverview}
\end{figure*}
%==================================================================
%==================================================================

HXR telescopes which focus light directly onto a small detector plane can achieve significantly lower non-solar background (higher sensitivity) than \rhessi, allowing for the identification of much fainter hot plasma sources. The \textit{Focusing Optics X-Ray Solar Imager} (\foxsi) sounding rockets have pioneered the development of direct-focusing telescopes optimized for solar observation, with science observations available from three successful flights\footnote{The fourth flight, \foxsi-4, is scheduled to occur in 2024.}. A quiescent AR observation made by \foxsi-1 significantly constrained the magnitude of a $>$8 MK component predicted by an EIS/XRT-only DEM of the region \citep{2014PASJ...66S..15I}. Additionally, nanoflare-associated hot plasma ($>$10 MK) was identified via \foxsi-2/XRT DEM analysis of another quiescent AR observed for 30 s during the $\sim$5 minute \foxsi-2 flight \citep{2017NatAs...1..771I}. \foxsi-2 also observed microflares, DEMs of which are presented in \cite{2020ApJ...891...78A}. These results (achieved with only a few minutes of observation time during each flight) demonstrate the power of this type of instrument to increase our understanding of the highest-temperature components of active regions.

The \textit{Nuclear Spectroscopic Telescope ARray} (\nustar) is an astrophysical focused HXR observatory, capable of solar observation with some operational caveats, discussed further in Section \ref{nuprep} \citep{Grefenstette_2016, 2013ApJ...770..103H}. Direct focusing allows \nustar\ to achieve a non-solar background more than three orders of magnitude lower than \rhessi\ \citep{Grefenstette_2016}. However, \nustar\ has limited detector throughput, meaning it records only a fraction of incident emission when observing bright solar sources (low livetime, see Section \ref{nuprep}). This limits our ability to utilize its tremendous sensitivity. Still, \nustar\ data has previously been used to place upper limits on the presence of hot plasma in one quiet AR \citep{Hannah_2016}. The properties of that region were found to correspond well to predictions by hydrodynamic simulations of low-frequency nanoflare ensembles \citep{Marsh_2018}. \nustar\ has also been used to study heating in two microflaring regions via DEM analysis \citep{Wright_2017, 2020ApJ...893L..40C}. 

To summarize past use of HXR observations to explore AR plasma heating: \rhessi\ and \foxsi\ studies have found intermittent evidence for a faint hot component at quiescent times, suggesting low-frequency heating, with the clearest detection presented in \cite{2017NatAs...1..771I} and confirmed via modeling in \citep{Marsh_2018}. \nustar\ has the capability to detect such a component as well, but has not yet been used for DEM studies of quiescent ARs. On the microflare side, only four events have been subject to DEM analysis with HXRs: \cite{Wright_2017} performed a DEM of the rise phase of an event, \cite{2020ApJ...893L..40C} performed pre-flare and flare-time DEMs, and \cite{2020ApJ...891...78A} performed DEMs of portions of two microflares observed during the $\sim$5 minute \foxsi-2 rocket flight. None of these have examined how the thermal distribution changes over the course of the flare.

In this work, we present a detailed thermal analysis of an AR observed by \nustar\ during an interval in which it both produced small microflares and also experienced times with no obvious transients. We perform DEMs utilizing \nustar, XRT (when available), and AIA for 170 distinct time intervals, allowing detailed inference into how the distribution of thermal plasma changes over time. This includes high-time-cadence DEM analysis of microflaring intervals.

We present this work as follows: in Section \ref{AR12712}, we introduce the AR under study, and discuss prior literature regarding this region. Section \ref{dems} discusses the DEM process: considerations regarding data preparation for each instrument, time interval selection, and the DEM method itself. Section \ref{res} summarizes and discusses results from the DEM analysis. Finally, Sections \ref{discuss} and \ref{conclu} provide further discussion and conclusions.

\section{Overview of Active Region Observations}\label{AR12712}

NOAA-designated AR 12712 was a bipolar region which produced two GOES C-class and $>$40 GOES B-class flares while transiting the solar disk. There was broad, multi-instrumental coverage of the region in the latter half of 2018 May 29, motivated by the flight of the \hic\ Sounding Rocket \citep{2019SoPh..294..174R}. This included $\sim$5 hours of coverage by \nustar, within a $\sim$7 hour period (15:55–23:20 UTC). A previous \nustar\ study of this active region focused specifically on flaring times \citep{duncan2021}, providing spectroscopy, time profile analysis, and imaging of seven events ranging from GOES sub-A1-class up to B1. That study found no evidence for a non-thermal accelerated particle distribution as a source of the observed \nustar\ emission in any of the microflares. Because of this, the analysis presented here presumes that all \nustar\ emission observed from the region has a thermal origin. 

\textit{Hinode}/EIS and XRT also observed this AR on 2018 May 29. While the EIS and \nustar\ observations overlap for just a few minutes, there are a number of intervals with joint \nustar/XRT coverage. The time profile of the AR over the entire observation interval is shown in Figure \ref{fig:lcoverview}, which includes \nustar\ livetime-corrected lightcurves in several energy ranges, as well as marked times during the \nustar\ coverage where XRT images are available. Figure \ref{fig:lcoverview} also shows the evolution of emission from the AR as observed by several AIA channels.

In conjunction with analysis of the \hic\ results, \cite{Warren_2020} performed a DEM analysis of the core of this active region using AIA, XRT, and EIS. The \hic\ flight and interval covered by the \cite{Warren_2020} DEM occurred during a \nustar\ night time. Discussion of the \cite{Warren_2020} results in comparison with those of this study is included in Section \ref{discuss}.

%==================================================================
%==================================================================
\begin{figure}
\includegraphics[width=0.47\textwidth]{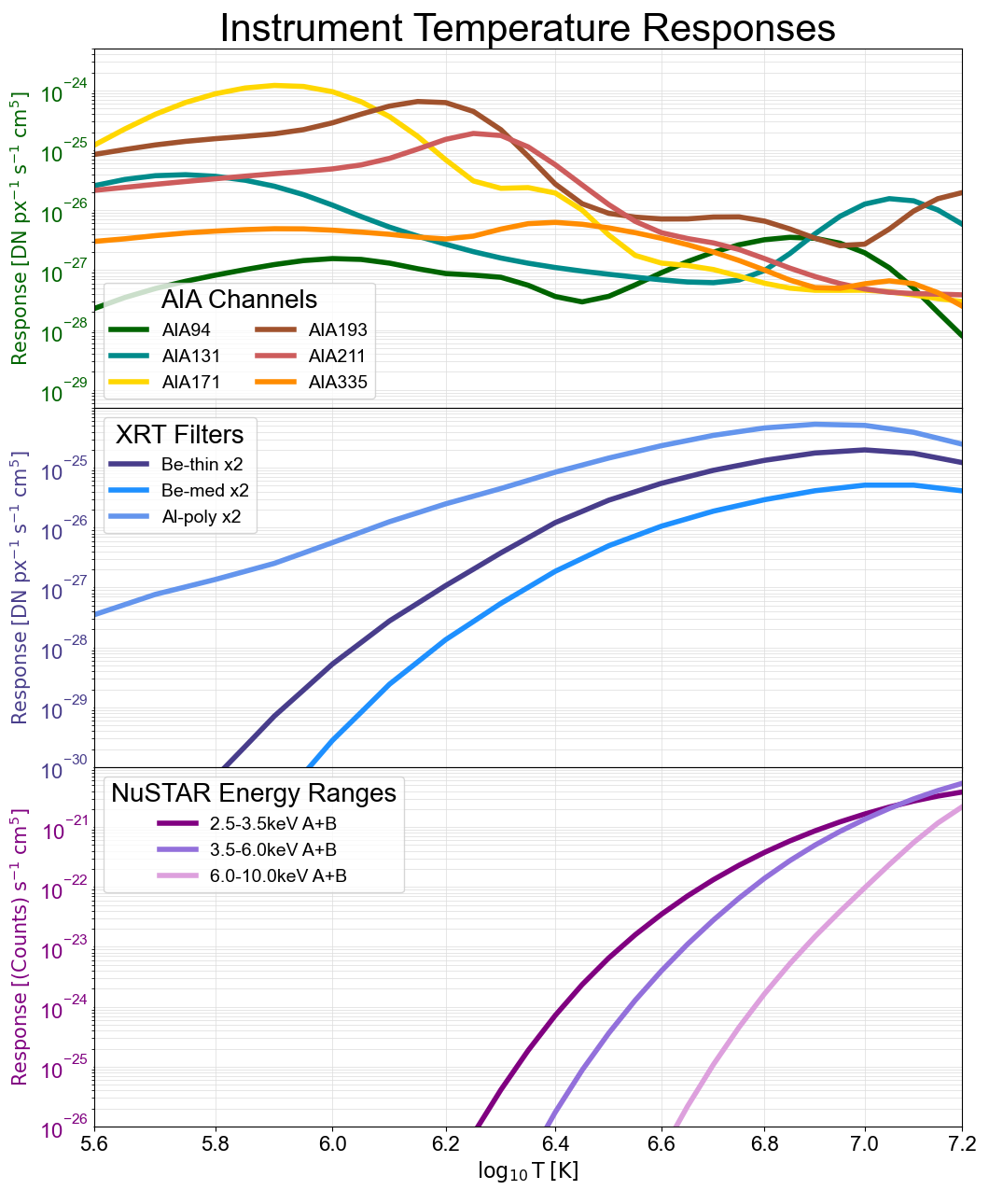}
\caption{Temperature responses of instruments used in DEM analysis. \textbf{Upper:} AIA. \textbf{Middle:} XRT. \textbf{Lower:} \nustar.}
\label{fig:allres}
\end{figure}
%==================================================================
%==================================================================

%==================================================================
%==================================================================
\begin{figure*}[!ht] 
\centering 
\includegraphics[width=0.85\textwidth]{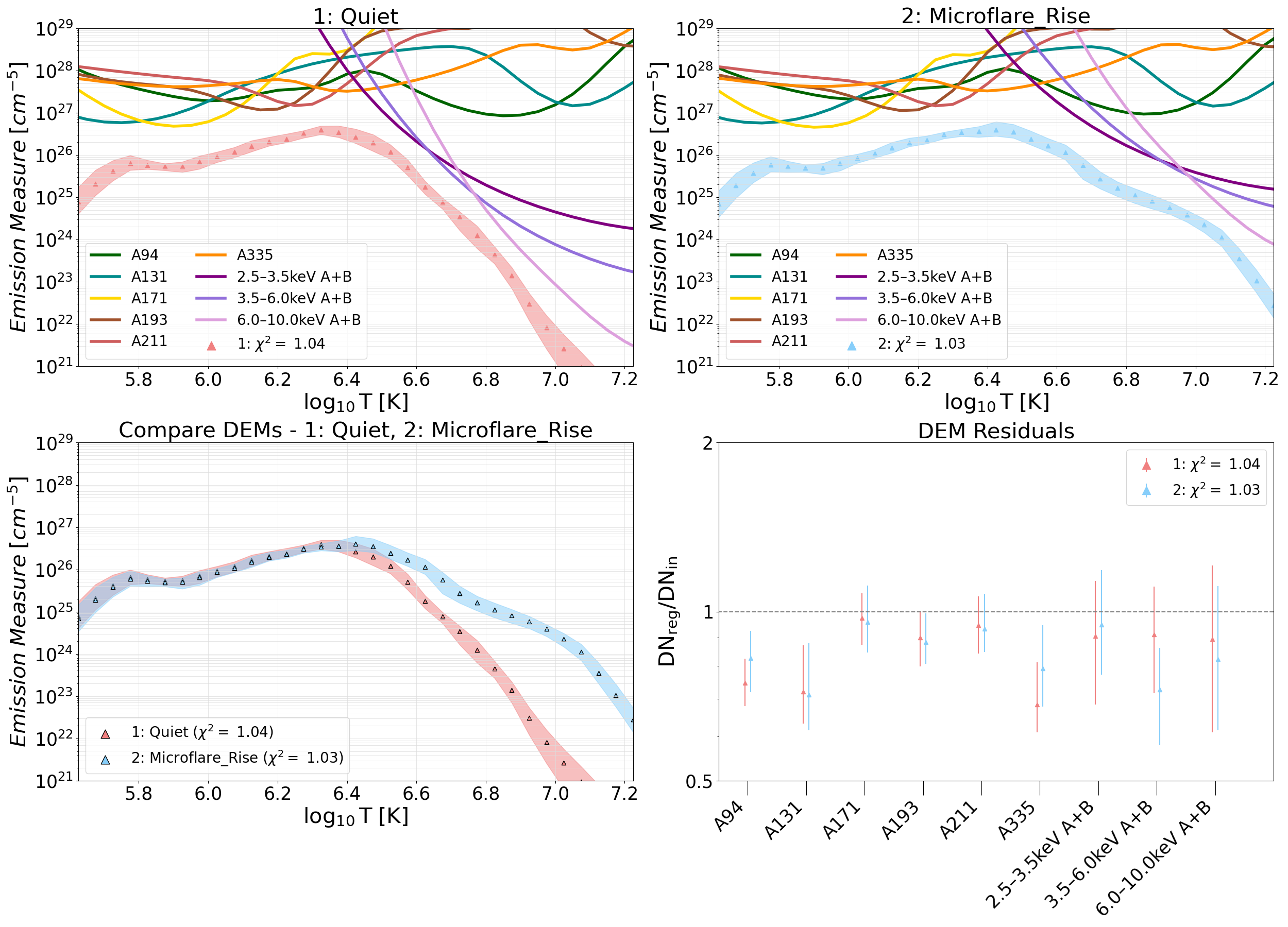} 
\caption{Results from DEM analysis of example intervals which represent extrema of the behavior of the region. EMD distributions are shown with shaded uncertainty regions giving the range of solutions found in all DEM iterations (see Section \ref{actualdem}). \textbf{Top Left:} A quiet time (20:51:30–20:54:30 UTC). \textbf{Top Right:} The impulsive phase of the largest microflare (16:22:15–16:22:45 UTC). Loci curves from each instrument involved are included in the top two panels. Both example DEMs are from times outside of the XRT data intervals, so no XRT data were used. \textbf{Lower Left:} Direct comparison. The quiescent interval (red) shows significantly less plasma at higher temperatures than is present during the impulsive phase of a microflare (blue). \textbf{Lower Right:} Residuals: DEM-predicted observations in each instrument, divided by the actual input. } 

\label{fig:DEMex}
 \end{figure*}
 %==================================================================
 %==================================================================
\section{DEM Analysis}\label{dems}

In the following, we introduce the concept of DEM analysis and details of the DEM calculation itself (Section \ref{actualdem}), the preparation of data and instrument responses for each instrument (Section \ref{prep}), and the time intervals selected for analysis (Section \ref{timeselect})\footnote{Code developed in this work for data preparation, organized execution of a large number of DEM calculations, and visualization of results has been made public at \url{https://github.com/jessiemcd/do-dem}}.

\subsection{DEM Calculation}\label{actualdem}

In a situation where multiple instruments have observed emission from the same thermal plasma source, the measurements made by each observing instrument ($M_{i}$) can be expressed,

\begin{equation}\label{DEM}
M_{i} = R_{iT}\times \xi(T),
\end{equation}

\noindent where $R_{iT}$ is a matrix containing the response of each instrument ($i$) as a function of temperature ($T$), and $\xi(T)$ is the line-of-sight DEM of the source as a function of $T$ (units: $[ \mathrm{cm}^{-5}  \mathrm{K}^{-1}]$). DEM calculation consists of inverting this expression in order to find $\xi(T)$, based on the $M_i$ and $R_{iT}$. The distribution of emission measure (EMD) as a function of temperature can also be extracted,

\begin{equation}\label{EMD}
EMD = \xi(T)dT
\end{equation}

\noindent (units: $[ \mathrm{cm}^{-5}]$). Figures in this work presenting DEM results all show the EMD distribution, as it is more straightforward to interpret with respect to results from HXR spectroscopy.

In this analysis, the DEM was computed using the regularized inversion method originally presented in \cite{hannah_kontar} for use with data from instruments aboard \textit{Hinode} and \textit{SDO} (referenced henceforth as the `DEMREG' method). Specifically, this study makes use of the Python implementation of DEMREG, which has been previously used with datasets involving \nustar\ \citep[e.g.,][]{2023SoPh..298...47P}. To estimate the uncertainty in each DEM solution, the DEM calculation was re-run iteratively (100x) while varying the input from each instrument within their input measurement uncertainties (described for each instrument in Section \ref{prep}). 

In addition to DEMREG, the standard \textit{Hinode}/XRT SolarSoft/SSWIDL procedure \texttt{xrt\_dem\_iterative2.pro}\footnote{This was the method used in \cite{2014PASJ...66S..15I, 2017NatAs...1..771I} and \cite{Ishikawa_2019}.} was used with the same input data and responses from all instruments in each DEM interval to generate alternative DEM solutions. This method also uses an iterative procedure to determine uncertainty bounds on the output solution. The addition of this secondary method (also iterated 100x) allows for confirmation that the physical conclusions resulting from this analysis are not found only with the use of one particular DEM method. 

The initial temperature range over which DEMs were computed was log(T) = 5.6–7.2 (see Section \ref{varange} for more discussion of temperature bounds). In order to compare DEM results to the input data, the DEM is convolved with the instrument temperature responses to generate DEM-predicted emission observed by each instrument. These residuals are shown in the lower-right-hand panels of all example figures, e.g. Figure \ref{fig:DEMex}. DEM residuals are discussed further in Section \ref{varange}.

\subsection{Data Preparation}\label{prep}

\subsubsection{AIA}\label{aiaprep}

\textit{SDO}/AIA provides continuous full-disk observation in a number of EUV passbands \citep{AIA_instrument}. Here, AIA images in the 94, 131, 171, 193, 211, and 335 \AA\ channels were used, as each have sensitivity to material at temperatures in the range associated with solar active region plasma. AIA channels are sensitive to multiple spectral lines with distinct peak formation temperatures; this leads to multiple peaks in each temperature response in Figure \ref{fig:allres}. 

For each image in each channel, level 1 data were converted to level 1.5 with the use of the \texttt{calibrate} subpackage within the \texttt{aiapy} Python package. A correction was made for the time-dependent degradation of the AIA instrument, also using \texttt{aiapy.calibrate}.

A spatial region was selected, encompassing the entire active region (a circular region, $150''$ in radius). The rate of emission observed across this region within each AIA image was expressed in units DN/s/pix. For DEM time intervals of 30 s duration, we used the AIA image taken nearest to the midpoint of the interval. For DEM time intervals of $>$30 s duration (see Section \ref{timeselect} for discussion of time interval selection), the AIA per-pixel rate was averaged across the interval, with images selected at a 30 s cadence. Uncertainties in each AIA channel were found using the \texttt{estimate\_error} function in \texttt{aiapy.calibrate}, to which a flat 10\% uncertainty was added in quadrature to account for assumed uncertainty in the AIA responses as was done in \cite{2023ApJ...943..180Z} to prepare AIA data for a joint DEM with \rhessi. The AIA temperature response for each channel was acquired using the routine \texttt{aia\_get\_response.pro} (from the SSWIDL \textit{SDO}/AIA library). The \texttt{hissw} Python package was used to incorporate the functionality of these and other SSWIDL libraries while performing analysis in Python.

\subsubsection{XRT}

\textit{Hinode}/XRT uses multiple combinations of filters to make spatially-resolved images of solar sources in the SXR band \citep{XRT_instrument}. XRT images of AR 12712 were available during only part of the \nustar\ interval, due to poor alignment between \textit{Hinode} and \nustar\ daylight times during the observing campaign (see Figure \ref{fig:lcoverview}). The XRT filter combinations used in this work were \textit{Al\_poly}, \textit{Be\_thin} and \textit{Be\_med}, each of which are sensitive to temperatures in the range between log(T) = 6.1–7.5 \citep{XRT_instrument}; their response in our range of interest is shown in Figure \ref{fig:allres}. XRT images are taken at a range of exposure times, in order to image sources across a wide range in brightness \citep{XRT_instrument}. For the two thinner filters (\textit{Al\_poly}, \textit{Be\_thin}), we selected only images with exposure times in specific ranges, in order to exclude saturated images and brief, low-resolution `flare patrol' images (\textit{Al\_poly}: between 0.1–1s, \textit{Be\_thin}: between 1–10 s).

For the selected images, the data were prepared by converting each image file from level 0 to level 1 using the \texttt{xrt\_prep.pro} routine from the \textit{Hinode}/XRT library within SolarSoft/SSWIDL. Pixel grade maps (an additional  \texttt{xrt\_prep.pro} output) were used to remove contamination spots (for \textit{Al\_poly} files only), as well as dust specks, hot pixels, and negative-value pixels for all files. After these steps, included pixels were selected using the same spatial regions as for AIA. To estimate uncertainties, $\xi$, in the individual pixel rates we used the expression, 

\begin{equation}
\xi = (1 + \sqrt{DN + 0.75} ) / (exposure\ time),
\end{equation} 

\noindent introduced to estimate XRT uncertainties for DEM analysis in \cite{2017ApJ...844....3L}, where DN represents the individual pixel values (units: $[DN]$). The total observed rate of XRT emission and uncertainty for each file (units: $[DN/s/px]$) were found by averaging the included pixel values and uncertainties. In the case where multiple suitable images were taken in the same XRT filter during the DEM time interval, the extracted pixel-averaged rates and uncertainties were averaged over all suitable images as well. Finally, the output uncertainties from this process were added in quadrature with an additional 10\% of the rate to account for uncertainty in the response. 

 The XRT temperature response for each filter combination was acquired using \texttt{make\_xrt\_temp\_resp.pro} (also from the \textit{Hinode}/XRT SSWIDL library). The default AIA emission model was used in the XRT response calculation, via the method described in the SolarSoft XRT Analysis Guide\footnote{https://xrt.cfa.harvard.edu/resources/documents/XAG/XAG.pdf}. A number of studies \citep[e.g.,][]{2011ApJ...728...30T, Wright_2017} have noted a discrepancy between emission measure distributions derived using XRT and those derived using other instruments. To resolve this we multiply the XRT temperature responses by a factor of 2, as has been done in prior studies involving \nustar\ \citep[e.g.,][]{Wright_2017, 2023SoPh..298...47P}. 

\subsubsection{\nustar\ }\label{nuprep}

\nustar\ consists of two co-aligned focusing X-ray telescopes, denoted hereafter by their focal plane modules, FPMA and B \citep{Grefenstette_2016, 2013ApJ...770..103H}. \nustar's telescopes have an angular resolution of $18''$ FWHM, or $58''$ HPD, significantly worse than the lower-energy instruments (AIA: $1.5''$, XRT: $2''$ ). For the purposes of this analysis, we consider the entire active region as one source, as the \nustar\ resolution limits our ability to spatially isolate emission from the AR core. 

Because \nustar\ is a spectrally resolved instrument, energy bin edges can be chosen to fit any scientific case at hand. Here, bins 2.5–3.5, 3.5–6, and 6–10 keV were chosen chosen to sample a range of the temperatures to which \nustar\ is sensitive. The temperature responses of these energy ranges are shown in Figure \ref{fig:allres}. The \nustar\ spatial region for each DEM interval was chosen by finding the center of mass of the \nustar\ emission during that interval, and selecting events recorded in a circular region $150''$ in radius around that point. 

The observed \nustar\ spectrum can be distorted by pileup. The effects of pileup are estimated by examining the relative incidence of different \nustar\ event ``grades", particularly the ``unphysical" grades that necessarily must involve more than one photon. In the time since the publication of \cite{duncan2021}, the \nustar\ heliophysics team has discovered that prior analysis methods underestimated the incidence of unphysical grade events, meaning that the pileup component of the \nustar\ spectra in that work was underestimated. Here, we find that a pileup correction \textit{is} necessary, and subtract an estimate of the pileup component from the \nustar\ spectrum before incorporation into DEM analysis\footnote{In terms of \nustar\ event grades: we select events of grades 0-4 for inclusion in DEM analysis, and correct for pileup by subtracting $1.25\times$ the observed spectrum in grade 21-24 events.}. Because this correction leads to a preferential reduction in the higher-energy part of the \nustar\ spectrum, we remain confident in the conclusion from \cite{duncan2021} that \nustar\ emission from the microflares observed in this AR has a solely thermal origin. 

During observations of active regions and flaring activity, the limited rate capability of \nustar's detectors means the percent of time the detectors spend processing events begins to dominate over the percent of time the instrument is ready to register a new photon (livetime) \citep{Grefenstette_2016}. \nustar\ livetime ranged from 0.2\%–4\% during this observation, which corresponds to an effective exposure time of around 5 minutes over the $\sim$5 hours of observation time. In preparing data for DEM analysis, the \nustar\ rate (in each region, time interval, and energy range) was corrected for instrument livetime to extract the expected true incident rate. The input uncertainties for each \nustar\ channel were found by adding 20\% of the observed rate in quadrature with statistical uncertainties based on the number of observed \nustar\ events.

In order to calculate the \nustar\ temperature responses, it is necessary to combine the expected X-ray spectrum from a thermal source with the \nustar\ instrument response in a given energy range. To find the former, a catalogue of simulated X-ray spectra for plasma of different temperatures was generated using \texttt{f\_vth.pro} (available in the SSWIDL XRAY library). This photon model includes both the thermal bremsstrahlung continuum, as well as spectral lines which appear in the energy range under consideration (2.5–10 keV). For the latter, the \nustar\ analysis pipeline\footnote{Specifically, the \texttt{nuproducts} module.} was used to generate spectral data products which contain information about the response of the instrument to incoming emission, including the energy-dependent effective area, detector response, and other factors. These are combined with the simulated thermal emission spectra as a function of temperature to generate a distinct temperature response function for every \nustar\ energy range. 

In order to take advantage of all information available, data and responses from both \nustar\ telescopes (FPMA, B) were summed in each energy range.

Prior \nustar\ studies involving spectroscopy of solar microflares have determined that there is a discrepancy in the \nustar\ gain that arises in the extremely high-count-rate (low-livetime) regime associated with observations of brighter solar sources. This discrepancy and a method used to correct for it in spectroscopy are presented in \cite{duncan2021}. In this analysis, it was not found to have an appreciable effect on DEM results, so no correction was applied to the \nustar\ inputs. The implications of the gain discrepancy for \nustar\ DEM studies are discussed further in Appendix \ref{gainagain}.

\subsection{Time Interval Selection}\label{timeselect}

We require at least 10 actual (not livetime-corrected) counts in \nustar\ in each energy range in each DEM time interval to achieve sufficient statistics for proper use of the DEMREG method. As discussed in Section \ref{nuprep}, limited \nustar\ livetime (range: 0.2\%–4\%, average: $\sim$2\% over this observation) means that the number of actual counts recorded by the instrument is far below the livetime-corrected estimate of the incident photon rate. 

In order to take advantage of better high-energy \nustar\ statistics at more active times while still integrating sufficiently long at quieter times, we calculate DEMs at an adaptive cadence. DEM intervals are at minimum 30 s in duration, and at maximum extend in duration however long it takes for the instrument to record 10 actual \nustar\ counts between 6–10 keV (sometimes several to $\sim$10 minutes). This results in 170 total time intervals over the $\sim$5 hours of \nustar\ observing time.

%==================================================================
%================================================================== 
\begin{figure*}[!ht] 
\centering 
\includegraphics[width=\textwidth]{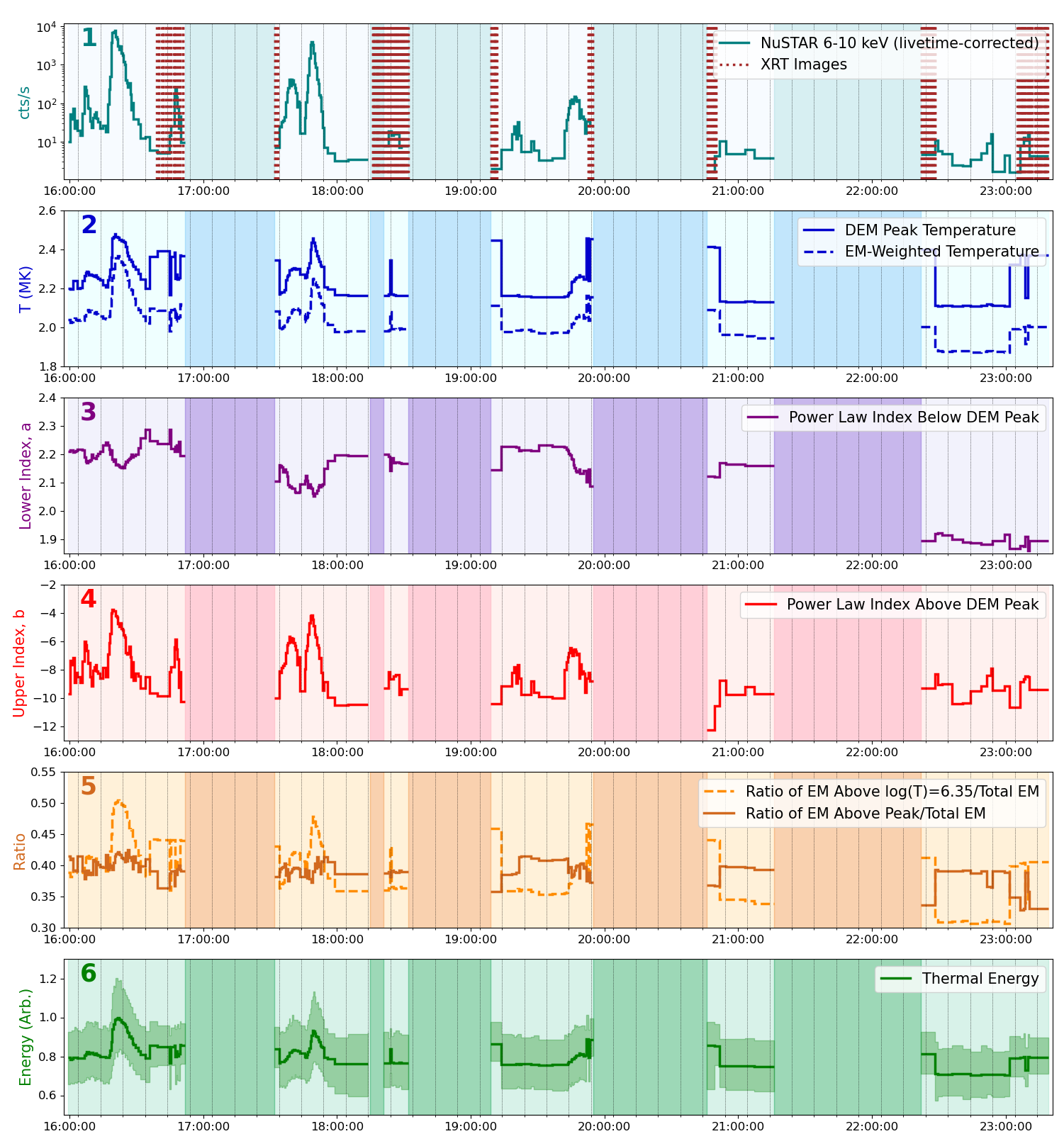} 
\caption{Evolution of the active region during the observation interval, as shown via parameters extracted from the resulting EMD at each time. Light and dark vertical shaded blocks indicate \nustar\ data availability, and dashed red lines in the top panel indicate times with both \nustar\ and XRT data available.  \textbf{1):} NuSTAR lightcurve (livetime corrected, 6–10 keV counts only), shown in log scale to emphasize much fainter transients in the later orbits as well as the larger microflares. \textbf{2):}  Temperature of EMD peak (found via Gaussian fitting), and emission-measure-weighted temperature over entire DEM interval (see Equation \ref{EMT}). For discussion of peak temperature (and other-parameter) discontinuities between XRT/no-XRT  times, see Appendix \ref{roles}. \textbf{3):}  Index of power law fit to EMD curve between log(T) = 6–6.35. \textbf{4):}  Index of power law fit to EMD curve above peak. \textbf{5):}  Total EM above the DEM peak temperature divided by total EM over entire temperature range, as well as total EM above a reference temperature (log(T) = 6.35) divided by total EM over entire temperature range. \textbf{6):} The evolution of the DEM-integrated thermal energy of the entire active region, shown in arbitrary units.} 
\label{fig:rainbow}
\end{figure*}
 %==================================================================
 %==================================================================
 
As AIA takes full-disk images in all of its EUV channels every $\sim$12 s \citep{AIA_instrument}, there are unique AIA data in every interval. During the intervals where XRT observed during the \nustar\ orbits, usable (non-saturated, full-resolution) images in each filter are available around every minute. However, due to the sparseness of the intervals with overlapping \nustar/XRT coverage (see Figure \ref{fig:lcoverview}), we do not make it a requirement to have a XRT image in every DEM interval. See Appendix \ref{roles} for discussion of the role of XRT in constraining the DEM when \nustar\ is also in use.

\section{Results}\label{res}

Figure \ref{fig:DEMex} shows the EMD resulting from DEMs evaluated during two time intervals: one where the AR was quiescent, and one during the impulsive phase of the largest observed microflare. Loci curves are shown (observed data divided by the instrument temperature-dependent response) for each of the instruments observing in the two time periods. Each of these curves forms a constraint on the DEM distribution. As expected, the microflare EMD shows significantly more plasma at higher temperatures than the quiescent example, while the distributions are more similar at lower temperatures. 

In the next subsections, we discuss the DEM results across \textit{all} time intervals. In Section \ref{res} we introduce a number of parameters extracted from the EMD and discuss their evolution throughout the observation interval. In Section \ref{microflare} we highlight the evolution of the EMD during a microflare. Finally, Section \ref{hott} discusses evidence for (and evolution of) higher-temperature components over time.

\subsection{Full-Observation DEM Evolution}\label{fullobs}

In order to understand the evolution of the AR EMDs throughout our observations, we extract a number of parameters describing aspects of the structure of these distributions. Figure \ref{fig:rainbow} summarizes the behavior of these parameters over the course of the full AR observation. Panel 1 of Figure \ref{fig:rainbow} shows the \nustar\ livetime-corrected 6–10 keV lightcurve.

\subsubsection{EMD Peak and Emission Measure Weighted Temperatures}
The peak of the EMD (Figure \ref{fig:rainbow}, Panel 2) is found between 2–2.6 MK consistently over the evolution of the active region. Increases in peak temperature are seen during larger transients, which heat enough plasma to shift the peak of the thermal distribution of the active region as a whole. This peak temperature is somewhat lower than that found in the bulk of prior AR DEM studies with EUV and/or SXR instruments (a range of studies find $\sim$3.2–4 MK \citep{2012ApJ...759..141W}). This result is expected, as this analysis includes a much broader spatial region (motivated by \nustar's limited spatial resolution, see Section \ref{nuprep}), rather than only a small portion of the AR core as is common practice when the instruments in use have higher spatial resolution \citep[e.g.][]{2016ApJ...824...56W}. The alternate DEM calculations using \texttt{xrt\_dem\_iterative2} found a similar range of peak values, as well as rising peak temperatures during flare times. However, the flare-associated increases were not as prominent with the use of this alternate method.

The emission-measure-weighted temperature ($T_{{EM}}$) is also shown in Panel 2, defined,

\begin{equation}\label{EMT}
T_{{EM}} = \frac{\int_{T}{{EMD}\times T}}{\int_{T}{{EMD}}}
\end{equation}

\noindent where the integrals may run over any range of temperatures where the DEM is defined (values shown in Figure \ref{fig:rainbow} were found via integration over the full temperature range). Like the peak, we see this value shift upwards during transients. In contrast to the behavior of the peak, the flare-associated enhancements in $T_{{EM}}$ are just as prominent when DEMs are calculated with \texttt{xrt\_dem\_iterative2}. 

Discontinuities in the time evolution of the peak temperatures are seen between times with and without XRT data available; see Appendix \ref{roles} for discussion of the effects of XRT on the DEM.

 \subsubsection{EMD Power Law Rise/Decay}

In addition to the two characteristic temperatures, we also examine the slopes of the EMD distribution above and below the temperature peak. We fit expressions of the form EM $\propto T^{a}$ (EM $\propto T^{-b}$) to the distribution below (above) the peak. In Figure \ref{fig:rainbow}, Panels 3 and 4 show the indices ($a$, $-b$, respectively) found from fits to the EMD in each time interval. 

The lower power law index ($a$) is related to the frequency of heating events occurring in active region loops, as well as the loop length \citep{2014ApJ...784...49C}. Here $a$ is fit between log(T) = 6–6.35 (1–$\sim$2.2 MK), and is found to range from $\sim$1.9–2.3 over the course of the observation, which is on the low end of the values found in prior EUV/SXR studies of AR cores ($2<a<5$, as summarized in Table 3 of \cite{2012ApJ...758...53B}, and which informed the analysis of \cite{2014ApJ...784...49C}). This result is consistent with the inclusion here of more lower-temperature material from outside the core. At flare times, $a$ dips downward, as heating of initially cooler material from sub-peak to super-peak temperatures leads to a more gradual slope up toward the peak. 

There are considerably lower values of $a$ seen during the final orbit. These low values occur because the DEM is strongly constrained by the 171\AA\ AIA channel in the lower part of the fit range (see Figure \ref{fig:allres}), and that channel sees a significant rise in emission during the final \nustar\ orbit (see Figure \ref{fig:lcoverview}) as a result of a transient that appears primarily in AIA 131 and 171\AA\ alone. 

Finding the power law index ($b$) of the decaying EMD above the peak provides a way to characterize the highest-temperature portion of the distribution, which is also crucial for investigating heating mechanisms. This index ranges widely during this observation (from $\sim$4–12), decreasing significantly during flaring intervals with more observed \nustar\ emission, as was also seen in the \foxsi-2 flare DEMs in \cite{2020ApJ...891...78A}. The evolution of the upper index closely follows the \nustar\ high energy (6–10 keV) lightcurve (as shown in Panel 1), because that measurement is the strongest constraint on the high-temperature DEM. The range of index values found here includes the range $7–10$ previously reported \citep{2012ApJ...759..141W, 2016ApJ...833..217B}. 
%==================================================================
%==================================================================
\begin{figure*}[!ht] 
\centering
\includegraphics[width=0.8\textwidth]{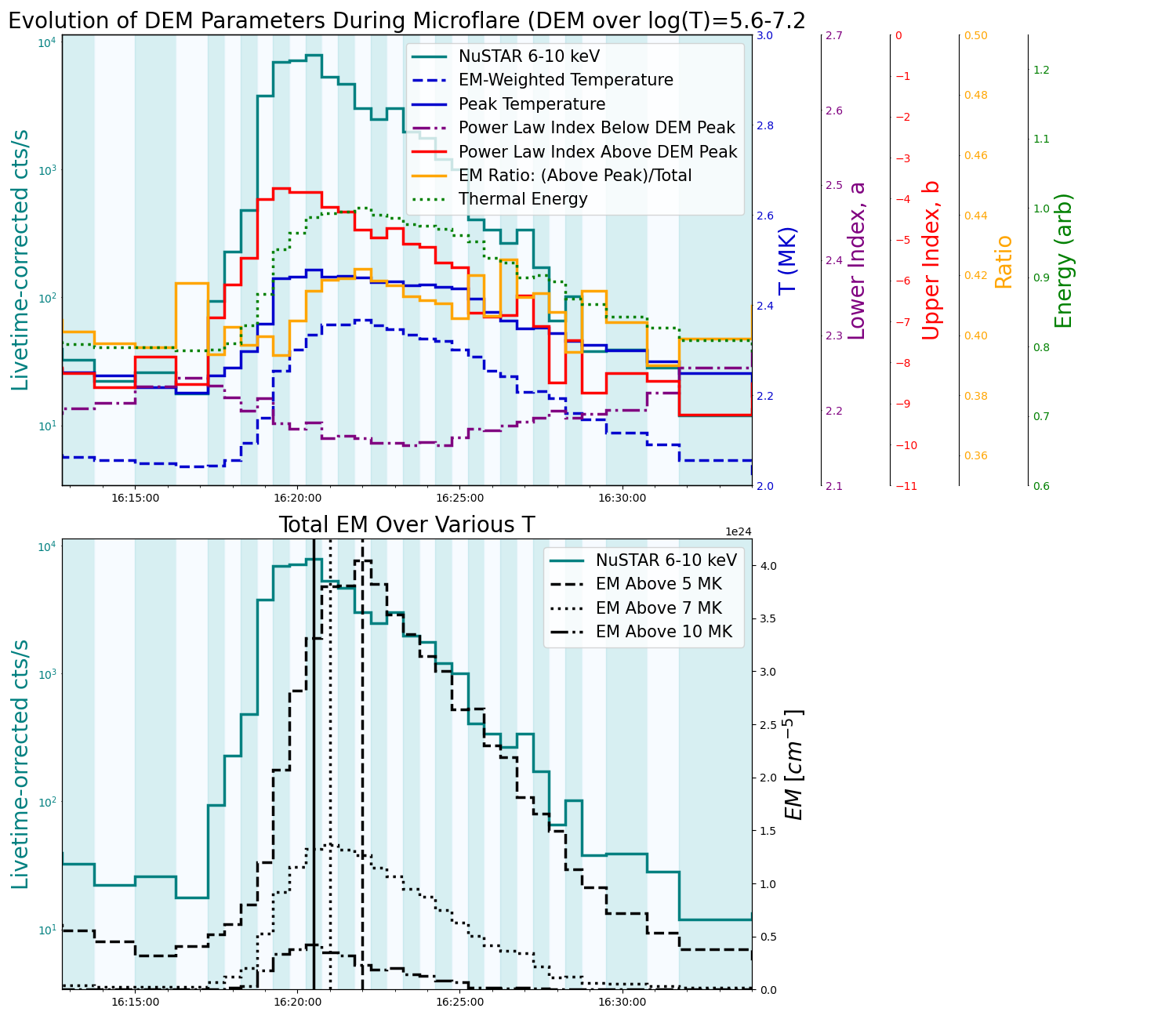}
\caption{Evolution of extracted parameters over a single microflare. Background light and dark shades are used to highlight the individual DEM time intervals in use. \textbf{Upper:} all parameters from Figure \ref{fig:rainbow} are shown over the flare interval, with several additional color-coded vertical axes defined to the right. \textbf{Lower:} Total emission measure (EMD-integrated) above three threshold temperatures are shown in conjunction with the \nustar\ 6–10 keV livetime-corrected lightcurve.}
\label{fig:microflare}
\end{figure*}
%==================================================================
%==================================================================
\subsubsection{Total Emission Measure Above/Below Peak}

The fifth panel in Figure \ref{fig:rainbow} shows the ratio of the total EM integrated above the peak and the total EM integrated over the full temperature range— i.e., the percent of material that is at temperatures above the peak as a function of time. We see that the higher-temperature material consistently makes up about 40\% of the total EM in the entire active region, with minor enhancements when more plasma is heated in a few of the larger flares. The peak itself shifts higher in temperature at flare times, so any flare-associated enhancement in this value indicates a significant increase in material at the highest temperatures (not just a shift in the distribution around the peak). For comparison, the \% of emission above a reference temperature ($\sim$2.2 MK) is also shown in this panel. This latter metric \textit{is} influenced by the shift in peak location. 

Discontinuities in the time evolution of these parameters are also seen at the beginning and end of the XRT times (see Appendix \ref{roles}). 
%==================================================================
%==================================================================
\begin{figure*}[!ht] 
\centering 
\includegraphics[width=0.9\textwidth]{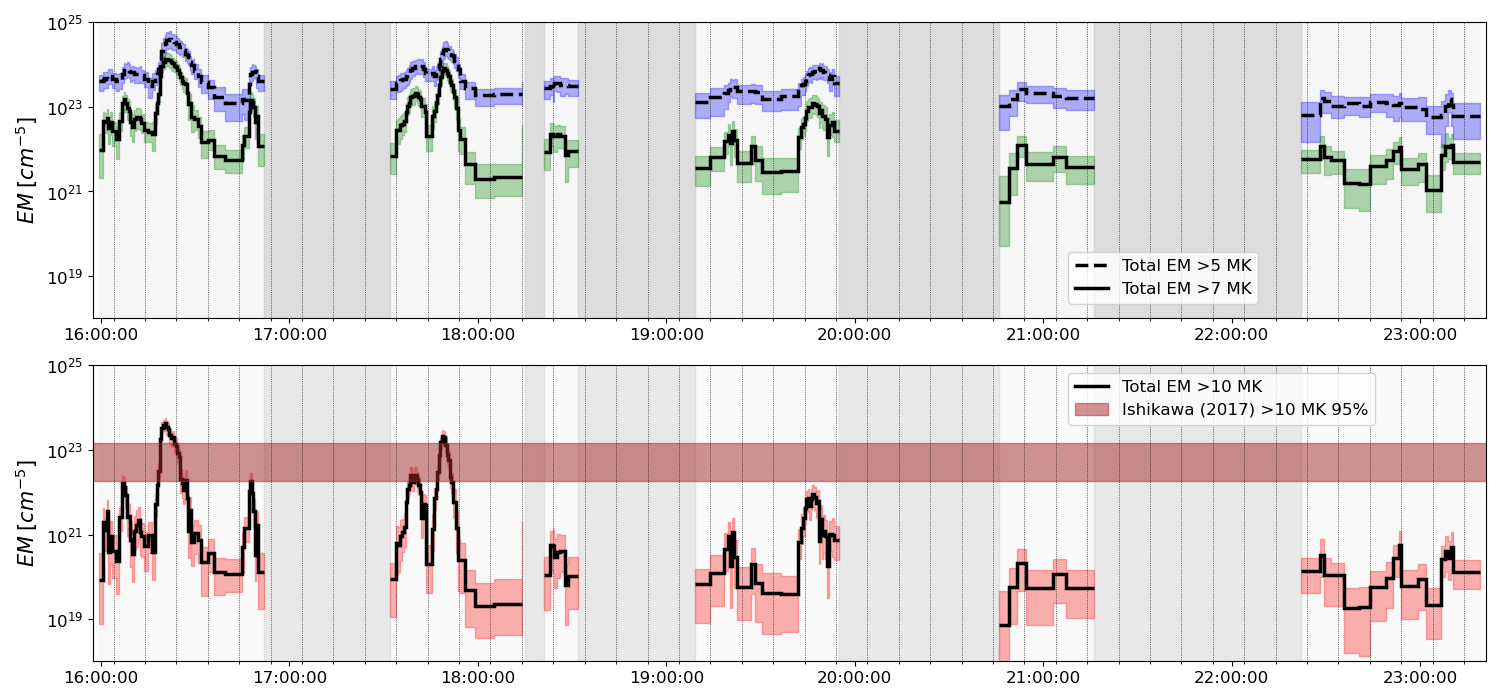} 
\caption{Total EM in the EMD as integrated above certain temperature thresholds, shown as a function of time. \textbf{Upper:} Total EM over 5 (7) MK, shown as dashed (solid) histogram with blue (green) uncertainty range. \textbf{Lower:} Total EM over 10 MK, also shown with uncertainty range (pink). Horizontal bar shows the estimated range of the amount of $>$10 MK plasma found via DEM analysis of a quiescent AR observed by XRT and \foxsi-2 \cite{2017NatAs...1..771I} (range: their DEM solutions with chi-squared values within 95\% occurrence probability ).}
\label{fig:totalEM}
 \end{figure*}
 %==================================================================
 %==================================================================
 
 %==================================================================
 %==================================================================
 \begin{figure*}[!ht] 
\centering 
\includegraphics[width=0.9\textwidth]{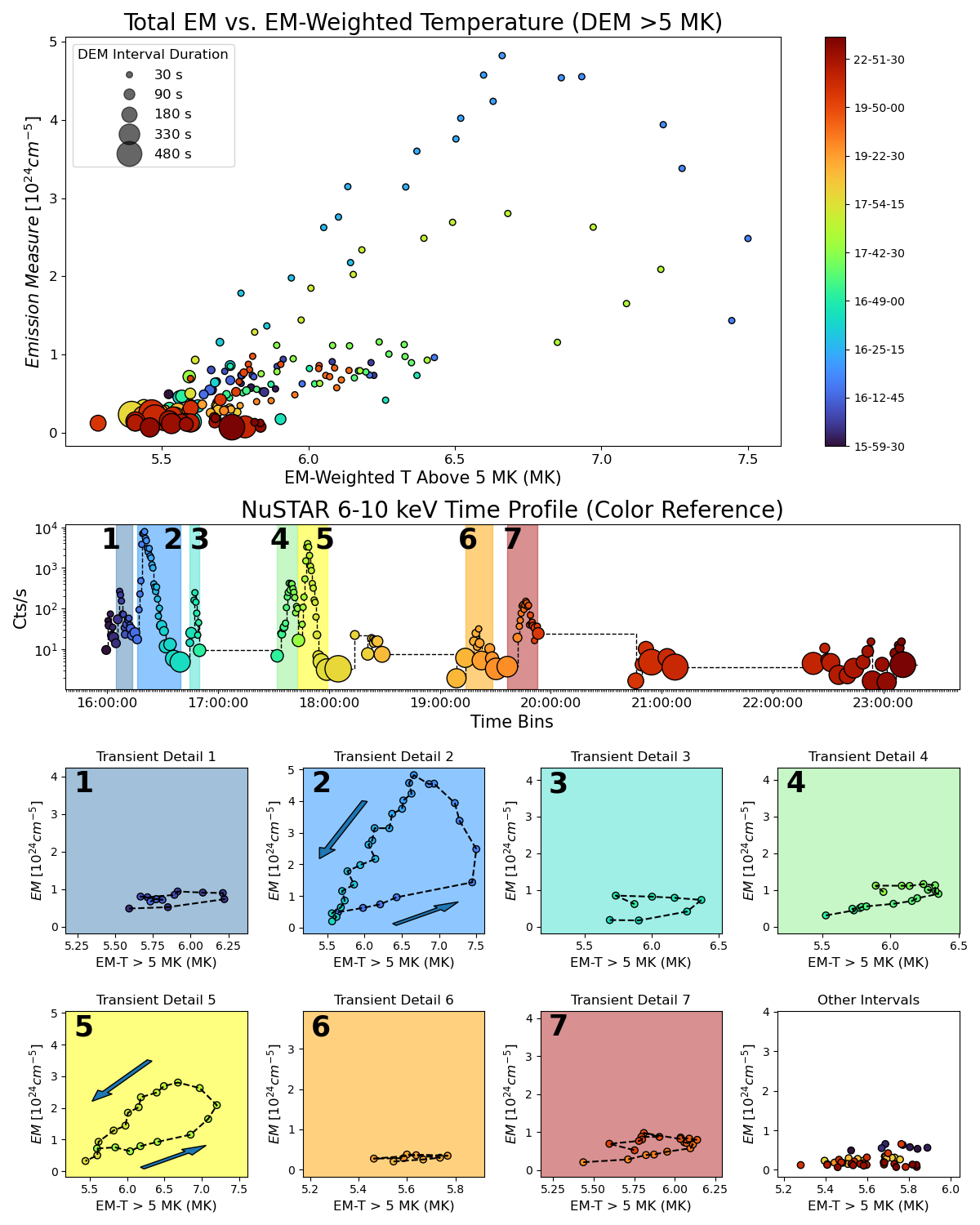} 
\caption{Evolution of the relationship between the total emission measure above 5 MK, and the EM-weighted temperature calculated above 5 MK (see Equation \ref{EMT}, and discussion in Section \ref{EMEMT}). Values from all DEM intervals are shown in the top panel, with transient intervals highlighted in the lower small panels. For the two largest transients (2, 5), arrows indicate the direction of the evolution in time, which starts and ends at the lower left. The smaller transients evolve in a similar counter-clockwise fashion. Colored points indicate time evolution. In the top plot, the size of each point indicates the duration of the DEM interval. The central panel shows the \nustar\ 6–10 keV rate during each DEM interval with the same point colors and sizes, for reference in interpreting the time evolution of the quantities above and below.} 
\label{fig:EM-EMT}
 \end{figure*}
 %==================================================================
 %==================================================================
\subsubsection{Thermal Energy}

The final panel in Figure \ref{fig:rainbow} represents the evolution of the thermal energy content of the active region. The thermal energy of an isothermal plasma volume is, 

\begin{equation}\label{thermal_real}
U_{{T}} = 3 k_B T \sqrt{EM fV}
\end{equation}

where $T$, $EM$ are the temperature and emission measure of the source, $V$ is the source volume, and $f$ is a filling factor. The DEMs in this analysis are performed over the active region as a whole, which contains many diverse plasma structures. This makes determination of the plasma volume and filling factor highly non-trivial. Additionally, it is desirable to estimate the energy of the full thermal distribution, rather than a single isothermal source. To do this, we define,

\begin{equation}\label{thermal}
U_{{T(arb)}} \propto \frac{\int_T \xi(T) \times T dT}{\sqrt{\int_T \xi(T) dT}},
\end{equation}

\noindent where $\xi(T)$ is the DEM as described in Equation \ref{DEM}, and we integrate over the temperature range used for the DEM calculation. Equation \ref{thermal} is a modified version of the expression for the thermal energy of a distribution of plasma given in \cite{2014ApJ...789..116I}. Here, factors $f$, $V$, and physical constants are omitted to instead give an estimate of thermal energy in arbitrary units, which can show the relative evolution of the energy content over time. 

Figure \ref{fig:rainbow} shows enhancement in this measure of the active region thermal energy at the times of every transient. It is fairly constant at other times, with the exception of discontinuities at the start of intervals where we have XRT data, which are discussed in Appendix \ref{roles}.

\subsection{Microflare Evolution}\label{microflare}

In Section \ref{fullobs}, we noted several EMD-extracted parameters that are modified at microflare times. Figure \ref{fig:microflare} shows the evolution of all parameters during a microflare. In the upper panel of Figure \ref{fig:microflare}, quantities from Figure \ref{fig:rainbow} are shown on the same axes. In the lower panel of Figure \ref{fig:microflare}, the total emission measure integrated above three threshold temperatures (5, 7, 10 MK) are shown, along with the \nustar\ 6–10 keV lightcurve.

Moving chronologically through the flare, the \nustar\ 6–10 keV count rate and upper power law index ($b$) are the first parameters to show flare-associated enhancement. This time period corresponds to an initial rise in the amount of plasma in the very highest temperatures. Next, we see the peak temperature, EM-weighted temperature, the thermal energy, and the EM curves in the lower panel begin to rise as more plasma is heated. Additionally, we see the lower power law index ($a$) decrease, which shows broadening of the EMD peak. 

The \nustar\ 6–10 keV lightcurve peaks before the EM curves, which then each peak successively in time (higher temperature thresholds peak earlier). In other words, the amount of plasma heated to the highest temperatures peaks before that heated to relatively lower temperatures. The \nustar\ 6–10 keV peak occurring before any of the EM curve peaks likely indicates the presence of significant plasma $>$10 MK in the rise/peak portion of this microflare (see additional evidence of this in Section \ref{varange}). 

We additionally see enhancement in the EM ratio during the flare, meaning that the total fraction of plasma above the EMD peak becomes an increased percentage of the total plasma distribution (by $\sim$4\%). This occurs despite the fact that the peak itself increases in temperature (which, on its own, would oppose an increase in this ratio). The EM ratio peaks latest of any of the extracted parameters shown, indicating that the total amount of plasma heated above the peak temperature continues to increase even as the high energy emission and hotter components are in decay. 

These results are all consistent with an initial energy release to heating of a small, hot volume, with later transfer of energy to more broad heating in the active region. In Section \ref{hott}, we will examine the evolution of the hot side of the distribution in more detail. 

%==================================================================
%==================================================================
\begin{figure*}[!ht] 
\centering 
\includegraphics[width=\textwidth]{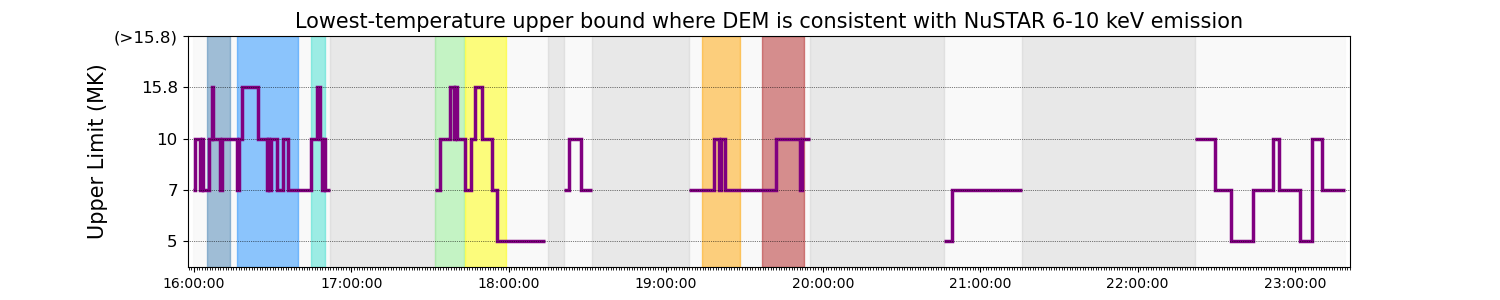} 
\caption{Results from varying the temperature range over which the DEM is calculated by lowering the upper limit. Restricting the highest temperature allowed in the DEM often causes inconsistency between the DEM-predicted and real input data in \nustar\ 6-10 keV. This figure shows the lowest upper-bound temperature at which the DEM-predicted 6–10 keV \nustar\ emission is consistent with the actual observed emission within uncertainty (see Section \ref{varange}). At the quietest times, a DEM over a temperature range that extends only up to 5 MK is consistent with observations, but during transients a broader temperature range is necessary for agreement. The same transients highlighted in Figure \ref{fig:EM-EMT} are highlighted here via multi-colored shading.} 

\label{fig:consist}
 \end{figure*}
 %==================================================================
 %==================================================================
\subsection{DEM Characterization of the Hottest ($>$5 MK) AR Plasma}\label{hott}

Figure \ref{fig:totalEM} shows the evolution of the DEM-estimated amount of plasma (in line-of-sight EM) above certain temperature thresholds as a function of time, with uncertainty ranges on the EM values from the iterative process described in Section \ref{actualdem}. The time evolution of the amount of material above these thresholds shows a close correspondence to the \nustar\ 6–10 keV lightcurve, as expected given the strength of that constraint in these temperature ranges. The amount of material above each threshold is enhanced both at identified microflare times, as well as in conjunction with smaller apparent X-ray transients. 

Looking in particular at the $>$10 MK EM, we compare the results to the estimated amount of $>$10 MK plasma found by \cite{2017NatAs...1..771I} in their \foxsi-2 study of AR 12234 during a quiescent interval. They estimate $1.8\times10^{22}–1.5\times10^{23}$ $\mathrm{cm}^{-5}$ above 10 MK. We see that AR 12712 possesses similar amounts of $>$10 MK plasma only during larger microflares, and that $\sim$3–4 orders of magnitude less $>$10 MK material is estimated to be present during quiet times or smaller transients ($\sim10^{19}$ $\mathrm{cm}^{-5}$; later results in Section \ref{varange} show us these data are actually consistent with \textit{no} $>$10 MK emission at quiet times). This order of difference was found using both the DEMREG and \texttt{xrt\_dem\_interative2} methods (DEMREG results shown). 

In the next two subsections, we examine the hottest material in this AR in more detail. Section \ref{EMEMT} traces out the time evolution of the relationship between the amount of $>$5 MK plasma and the emission-measure-weighted temperature, and Section \ref{varange} varies the temperature range used for the DEM calculation in each interval, to investigate whether the inclusion of material above certain thresholds is necessary for a good DEM solution. 

\subsubsection{Total Emission Measure vs. Emission-Measure-Weighted Temperature}\label{EMEMT}

 Here, we consider the evolution of the $>$5 MK material in the AR, particularly examining the interplay between how much plasma is observed above that threshold (total EM, found by integrating the EMD distribution from $\sim$5.0–15.8 MK), and the emission-measure-weighted temperature above that threshold ($T_{{EM}}$, found by evaluating Equation \ref{EMT} with both sums taken over that same range). Figure \ref{fig:EM-EMT} illustrates the motion of the distribution through the phase space defined by these two quantities as a function of time. 

The top panel of Figure \ref{fig:EM-EMT} shows these values for all 170 DEM intervals. Points are colored with respect to their order in time, with a corresponding time-stamped colorbar and a \nustar\ lightcurve serving as a key in the panel immediately below. Additionally, the duration of each DEM interval is expressed via the size of the markers, with longer-duration DEMs (in quieter times, see Section \ref{timeselect}) appearing as comparatively larger circles. We see that points from quiet times are grouped toward the bottom left, indicating relatively small amounts of plasma above 5 MK, as well as a low emission-measure-weighted temperature. At times of transients, there are excursions up and to the right. 

To highlight transient behavior, the lower set of panels shows the time evolution during specific intervals (highlighted and numbered in the lightcurve). To allow comparison of the shape of the evolution between transients, the aspect ratio of these plots is held fixed to that of the top plot (though the plot limits change to better highlight the behavior of transients of different magnitude). For the larger transients (labels 2 and 5, B- and mid-A-class microflares, respectively), arrows indicate the counterclockwise path the EMD follows through this space, starting at the lower left.

Examining the larger events, we see that the $>$5 MK portion of the EMD shows an initial sharp increase in $T_{{EM}}$ with a smaller simultaneous increase in EM. This change indicates that \textit{the initial heating occurring in the microflares is heating of material that is already relatively hot}. Later in the evolution of these events, there is a much sharper rise in EM while $T_{{EM}}$ actually decreases. This suggests later heating of much more material that was initially cooler, consistent with a picture involving a rise in density in coronal loops later in the flare evolution, once flare-heated chromospheric material evaporates upward. 

During quiet times and smaller transients (smaller- and sub-GOES A-class activity), we see initial increases predominantly along the $T_{{EM}}$ axis, with lesser enhancement of total EM, followed by a corresponding decay back to the pre-flare condition. These smaller transients repeatedly trace out similarly-shaped diagonal paths in this space, indicating that over the course of this observation, plasma heating follows a similar process in events across a range of scales.

 \subsubsection{Variable Temperature Range DEMs}\label{varange}

The temperature range over which the DEM is calculated in the bulk of this analysis is log(T) = 5.6–7.2, or $\sim$0.4–15.8 MK. The lower bound was chosen to include the peaks of the response of all the included AIA channels (see Figure \ref{fig:allres}). The upper bound was chosen to extend a bit above the temperatures often seen in \nustar\ spectroscopy of similar-magnitude transients (often around 10 MK). We are interested to explore whether allowing plasma at temperatures up to ~15.8 MK is really necessary for a result that is in good agreement with the inputs, particularly the 6–10 keV \nustar\ emission, which provides the most rigorous high-temperature constraint. To examine this, we re-run DEM analysis for every interval using three additional upper bounds on temperature: $\sim$5, $\sim$7, and 10 MK, respectively. For each new solution, we record whether the DEM-predicted data is consistent with the actual \nustar\ input data in the 6–10 keV bin within uncertainty. 

Before completing these variable temperature range DEMs, we note a property of the existing results that has the potential to affect this analysis: across time intervals, there is a tendency for DEM residuals to trend, on average, slightly below 1 (visible, for example, in both cases presented in Figure \ref{fig:DEMex}). This indicates that this method results in DEM-predicted emission of slightly lower magnitude than the actual observations, equivalent to a slight underestimate in the amount of material present across the temperature range. 

A constant factor $f = \frac{1}{n}\times \sum_r \frac{1}{r}$ can be found from the $n$ original residuals, $r$, that can be multiplied by the EMD to find a new solution with DEM-predicted residuals more uniformly distributed around 1. Comparing results with and without this EMD scaling factor across all prior sections of the paper, we see no significant effect on the results. This is because the rescaling is constant across temperature (so does not effect the DEM shape), and is of small magnitude with respect to the EMD ($f$ is found to be always $1 < f < 2$). However, we do find the application of this correction factor necessary for the variable temperature range DEM analysis, as the rescaling has a more significant effect on the residuals themselves (by design), and we are directly utilizing the \nustar\ 6-10 keV residual to determine if we see evidence for material above a given temperature range.

In Figure \ref{fig:consist} we show the lowest-temperature upper bound where the DEM-predicted data is consistent with the \nustar\ 6–10 keV input within uncertainty, as a function of time. At quieter times, a DEM over a range extending only up to 5 MK might produce a consistent result, but during more active intervals allowing plasma up to higher temperatures is clearly needed. Using the DEMREG method, we find that allowing material above 10 MK is necessary in 5/7 of the marked microflares (colored shading). Using \texttt{xrt\_dem\_iterative2}, it is necessary in the 2/7 largest events. 

The higher-temperature spectral components found via spectroscopy of these microflares in \cite{duncan2021} ranged from 7.8–10.1 MK (with all but one $<$10 MK). These results are not inconsistent, but rather reflect the fact that spectroscopy of a thermal plasma distribution using isothermal models extracts temperatures that are best viewed as characteristic temperatures representing what is in fact a complex multi-thermal distribution. DEM analysis provides more detailed information about the thermal distribution, and suggests that in each of these events there is a notable extension above 10 MK.

We note that even outside of the confirmed microflaring times (during quiescent times, and small x-ray transients that are much too faint to be seen in GOES) there are intervals where plasma above 5 or 7 MK is needed for a solution consistent with the \nustar\ observations. However, there are no intervals outside of microflares where $>$10 MK material is necessary. This, again, contrasts greatly with the results of \cite{2017NatAs...1..771I}, where a significant amount of $>$10 MK material was identified in a different quiescent AR.

\section{Discussion}\label{discuss}

The objective of this analysis was to examine the thermal distribution of plasma in AR 12712 through both flaring and non-flaring times, via time-resolved differential emission measure analysis. We first discuss our results in conjunction with prior literature involving quiescent AR DEMs. 

The range of values found for the upper power law ($b$) fit to the EMD decay above the peak intersected with the range of values found in other works, as summarized in \cite{2016ApJ...829...31B}. We note that the second paper in that series \citep{2016ApJ...833..217B} includes discussion of the challenges of meaningful interpretation of this index, due to departure of the EMD from a power law relationship. This departure leads to high sensitivity of the fit to the chosen range of temperatures over which it is calculated. We observed similar issues in this analysis.

Several parameters describing the EMD distribution (peak temperature, rise index $a$) deviate from established literature using EUV and SXR instruments. The EMD peak was found consistently between 2–2.6 MK over the observation interval, distinct from both the $\sim$3.2–4 MK found in a range of prior EUV/SXR studies of other regions \citep{2012ApJ...759..141W}, as well as the 4 MK peak found by \cite{Warren_2020} for \textit{this} AR using EIS, XRT and AIA. Similarly, $a$ was found to range between 1.9–2.3 over the course of the observation (compare to the range 2 $<$ a $<$ 5 presented in \cite{2012ApJ...758...53B}). 

These differences are consistent with the fact that a larger portion of lower-temperature material in the AR is included in our DEMs, rather than just the AR core. This lower-temperature material is a consequence of the large spatial regions chosen for analysis, motivated by the comparatively poor spatial resolution of \nustar. This limitation is a consequence of \nustar's specific design; improved resolution (5'' FWHM) has been achieved with \foxsi-3 and even higher-resolution focused HXR instruments are currently in development \citep{10.1117/12.2594701}. 

Due to the highly distinct magnetic environment of the AR core with respect to the outskirts of the AR or surrounding quieter regions, it is reasonable to assume that heating processes may proceed very differently between these coronal features. This strongly motivates the implementation of future focused HXR instruments that combine high sensitivity with improved spatial resolution, allowing independent measurement of the thermal distribution of the AR core (as opposed to the combination of the core and its surroundings).

Further considering the hot side of the EMD, we found significantly less $>$10 MK material in this region at quiescent times than was found in the quiescent AR analyzed by \cite{2017NatAs...1..771I} ($\sim10^{19}$ vs. $>10^{22} \mathrm{cm}^{-5}$). The larger region used in this analysis (by a factor of $\sim$8: here, circular with radius $150''$, vs. their $100''$ square region) contributes to this discrepancy: since more hot plasma is expected to be located in the AR core, the use of a larger region including parts of the AR with little high-temperature material has the effect of depressing the line-of-sight emission measure at high temperatures (essentially, a spatial average of the emission measure across the entire region under consideration). However, this cannot fully explain the difference. Estimating that the AR core takes up roughly 10\% of the area in our region, and assuming that \textit{all} material $>$10 MK exists in the AR core alone, there would only be a difference of one order of magnitude in the amount of $>$10 MK plasma found in our region vs. a hypothetical ``AR core only'' DEM analysis using a higher-resolution HXR instrument. If $>$10 MK material exists outside the AR core, the difference would be even smaller. This suggests that the distribution of thermal plasma in this region is fundamentally different than the region observed by \cite{2017NatAs...1..771I}. 

In \cite{Marsh_2018}, the observed properties of the AR studied by \cite{2017NatAs...1..771I} were found to $>$99\% confidence to be consistent only with models involving low-frequency (nanoflare) heating, rather than high-frequency heating. Our AR was subject to similar modeling efforts in \cite{Warren_2020}, which concluded that high-frequency heating was most consistent with their observations (EIS, XRT, and AIA observations made during 18:56–19:01 UT, between two \nustar\ orbits). The significantly lower amount of $>$10 MK plasma found in this analysis in comparison to \cite{2017NatAs...1..771I} makes our results qualitatively consistent with that conclusion, but a rigorous determination would require a comparison between modeling results and the DEM derived using the \nustar\ HXR observations to firmly constrain the higher-temperature material. 

We additionally note the potential for the combination of similar modeling methods with observations made by future focused HXR instruments with higher spatial resolution: achieving a HXR-constrained DEM of the AR core in combination with modeling would allow us to both infer properties of the heating processes occurring, and evaluate whether the heating we see is sufficient to sustain the observed thermal distribution over time. 

In examining microflare evolution, we confirm a picture wherein the event starts with the heating of a small amount of plasma to very high temperatures, followed by transfer of thermal energy to later heating of a larger amount of material to lower (yet still elevated) temperatures. This is in agreement with results from time profile analysis in \cite{duncan2021}. Adding to this picture, the analysis of Section \ref{EMEMT} particularly suggests that the initial plasma heating to the highest-achieved temperatures is dominated by heating of plasma that is \textit{already $>$5MK}. This analysis is consistent with initial energy release occurring in the hotter AR core. Later in each event, as we see a sharp rise in EM with a steady or decreasing $T_{{EM}}$, we infer that cooler material from the chromosphere is being heated and evaporating upward, increasing the density of loops in the AR core. 

Strikingly, the flare-time evolution of the plasma distribution in EM vs. $T_{{EM}}$ space in Figure \ref{fig:EM-EMT} has close similarities to the evolution of the plasma density as a function of temperature in prior EBTEL simulations of plasma heating in a single AR loop (compare with Figure 4 in \cite{2016ApJ...829...31B}). Like larger flares, microflares at this scale are expected to involve many coronal loops (rather than a single loop structure); it is intriguing that the behavior of a source consisting of large ensemble of loops seems similar to that predicted for just one. This strongly motivates future studies involving both DEMs and hydrodynamic modeling of the same events, to better understand how energy release and heating occur over a range of scales.

\section{conclusions}\label{conclu}

In this analysis, we have presented the first HXR-constrained, time-resolved DEM analysis of an evolving active region. At microflare times, we have shown a detailed picture of the thermal evolution, involving initial heating of already-hot plasma in the AR core to very high temperatures followed by later broad heating of surrounding material to cooler (yet still elevated) temperatures. At non-flaring times, we estimate that significantly less ($>$3 orders of magnitude) plasma above $10$ MK exists in this region than was seen in a prior HXR quiescent AR DEM study \citep{2017NatAs...1..771I}. The significant differences between these results strongly motivate further study of additional active regions with HXR coverage at quiescent times, as it is not presently clear which (if either) of these results is typical of solar active regions as a whole. Progress toward this goal will be possible via additional existing observations made by \nustar\ (some $\sim$100 hours of observations made in campaigns between 2014–2023). A solar-dedicated focused HXR observatory capable of making these observations near-continuously (and without \nustar's limited livetime and spatial resolution) would allow conclusive determination of the characteristic thermal structure of solar active regions. In combination with observation-informed modeling efforts, these observations would clarify the mechanisms responsible for AR heating outside of large transients.

\begin{acknowledgments}
Jessie Duncan’s research was supported by an appointment to the NASA Postdoctoral Program at the NASA Goddard Space Flight Center, administered by Oak Ridge Associated Universities under contract with NASA. Work at the University of Minnesota was supported under an NSF CAREER award (AGS 1752268). \nustar\ observations were completed as part of the NASA \nustar\  Guest Observer program (80NSSC18K1744). This research made use of the \nustar\ Data Analysis Software (NUSTAR-DAS) jointly developed by the ASI Science Data Center (ASDC, Italy) and the California Institute of Technology (USA).
\end{acknowledgments}


\vspace{10mm}
\facilities{\nustar, \textit{Hinode}, \textit{SDO}}


\software{aiapy \citep{Barnes2020}, 
		sunpy \citep{sunpy_community2020},
		astropy \citep{2022ApJ...935..167A},
		scipy \citep{2020SciPy-NMeth},
		numpy \citep{harris2020array},
		matplotlib \citep{Hunter:2007},
         hissw \citep{will_barnes_2022_6640421}, 
         do-dem \url{https://github.com/jessiemcd/do-dem},
         SSWIDL \citep{1998SoPh..182..497F}
         }



\appendix

\section{Comparative Roles of \nustar, XRT, and AIA in Constraining the DEM}\label{roles}

\begin{figure*}[!ht] 
\centering 
\includegraphics[width=0.85\textwidth]{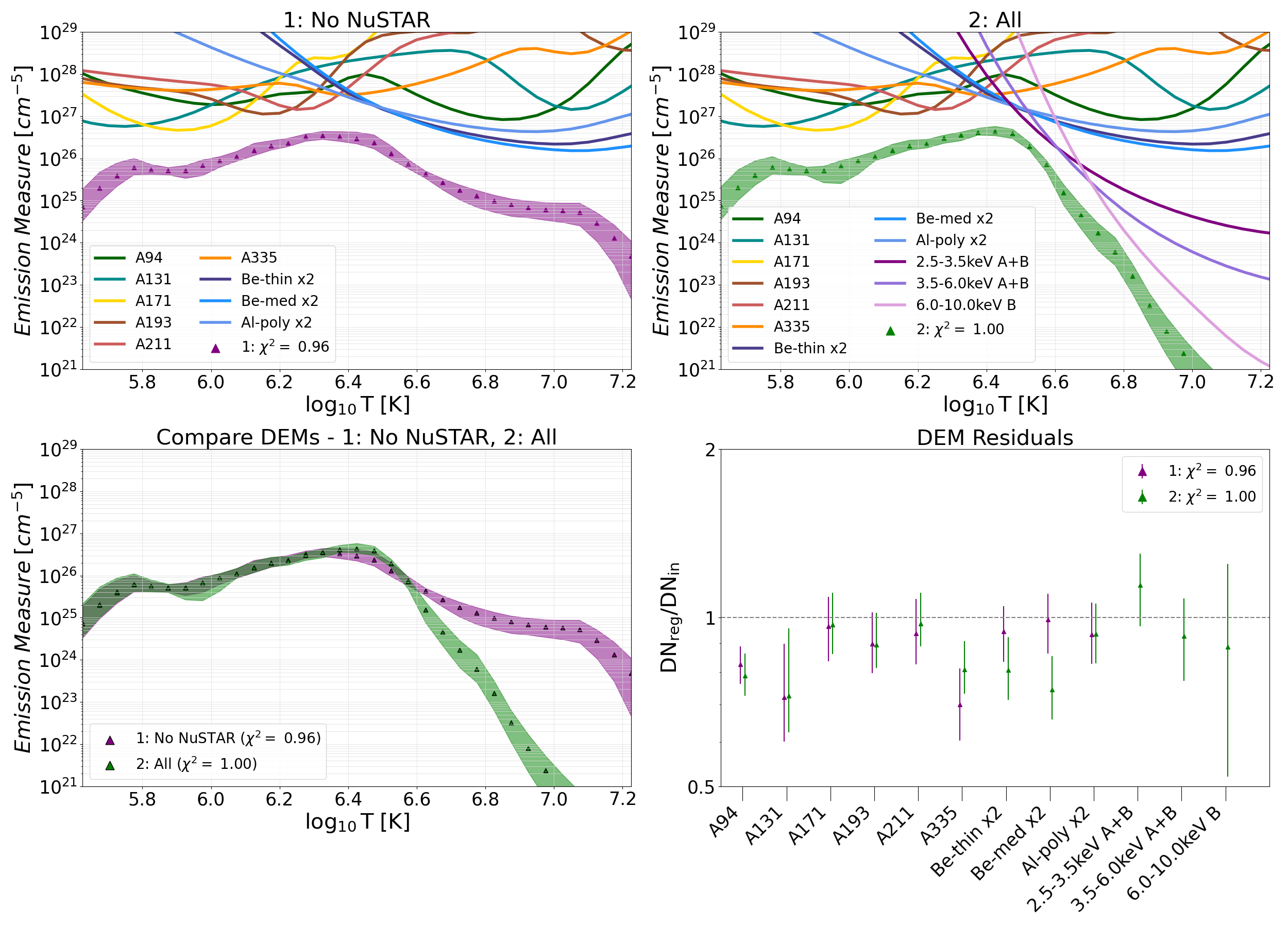} 
\caption{Comparison of DEM results during a quiescent interval (19:09:00–19:13:45 UTC) with and without the addition of \nustar. High-temperature behavior of the EMD is significantly modified by the \nustar\ constraints (see discussion in Appendix \ref{roles}).} 
\label{fig:nonu}
 \end{figure*}

Several studies have demonstrated that DEMs using AIA and XRT alone are prone to over-estimate the amount of plasma above 5 MK (for example, \cite{ 2009ApJ...704..863S}, \cite{2023SoPh..298...47P}, as well as the differing results found by \cite{2020ApJ...891...78A} and \cite{2016ApJ...833..182S} for the same region). We replicate this result: when \nustar\ constraints are not used and the DEM analysis is performed with AIA and XRT alone, the total emission measure above 5 MK is an order of magnitude higher than the result when using \nustar. This discrepancy is even more striking when isolating even higher-temperature components: the AIA/XRT-only EMD shows 2 (3) orders of magnitude more plasma above 7 MK, and 3 (4) orders of magnitude more plasma above 10 MK at a flaring (quiescent) time. Figure \ref{fig:nonu} shows results with and without \nustar\ during a quiescent interval (19:09:00 to 19:13:45 UTC), clearly displaying the differing behavior at high temperatures. 

\begin{figure*}[!ht] 
\centering 
\includegraphics[width=0.85\textwidth]{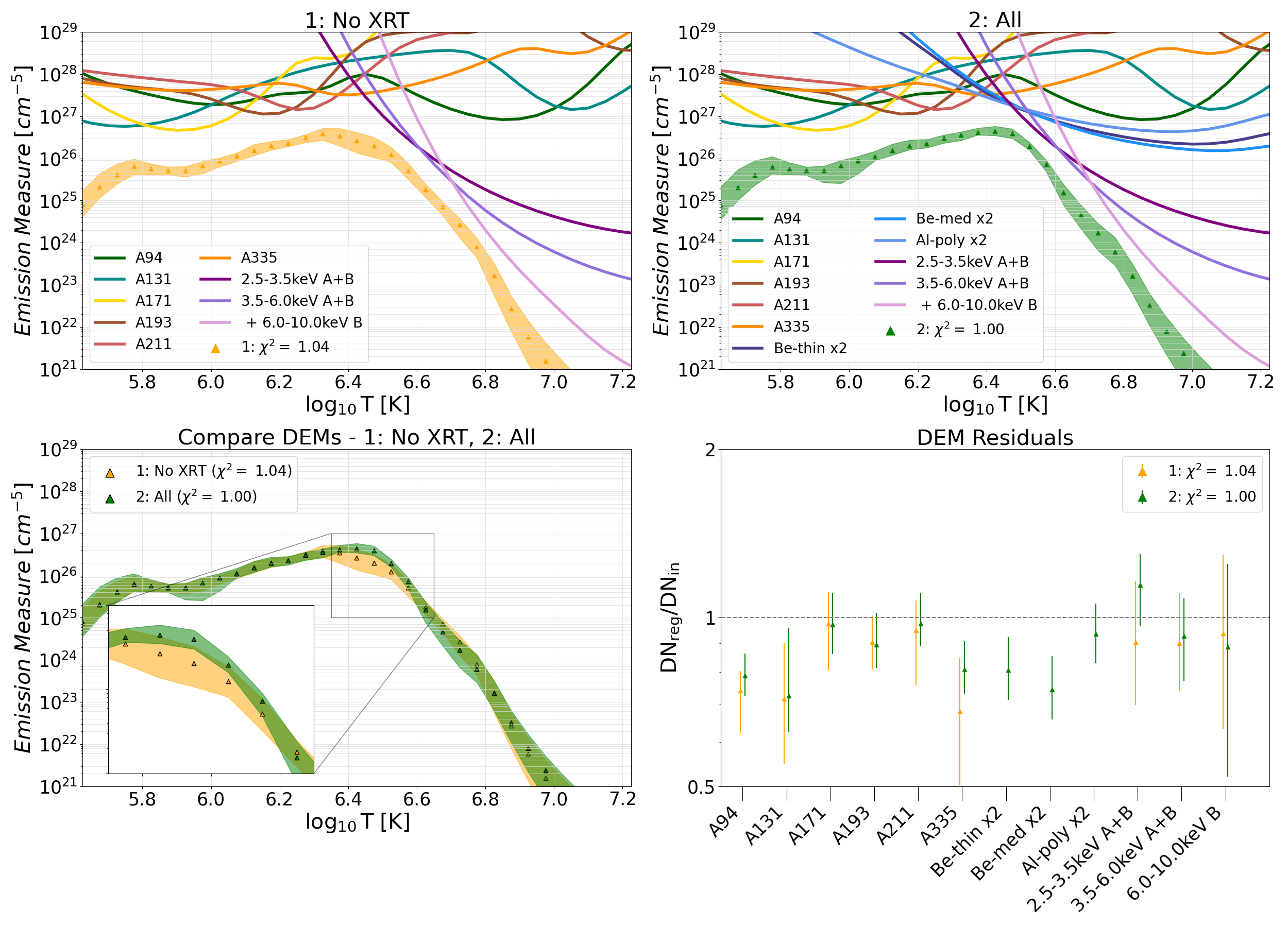} 
\caption{Comparison of DEM results during a quiescent interval (19:09:00–19:13:45 UTC) with and without the addition of XRT. While the EMDs are consistent with each other within their uncertainties over the full interval, the addition of XRT changes the solution slightly in the highest-EM region, resulting in noticeable differences in peak-associated parameters (see discussion in Appendix \ref{roles}, and Figure \ref{fig:rainbow}). } 
\label{fig:noxrt}
 \end{figure*}

Examining the EM loci curves in Figures \ref{fig:nonu} and \ref{fig:noxrt}, it is clear that the AIA channels provide the strongest constraints at low temperatures, and the \nustar\ energy ranges at the highest temperatures. However, there is a narrow window in temperature near the middle of the range (between log(T) = $\sim$6.4–6.5) where the most stringent constraint is provided by the XRT SXR filter combinations. Interestingly, this temperature is in close proximity to the peak of the EMD, and as such small changes to the constraints in this temperature range can have noticeable effects on certain output parameters. 

%==================================================================
%==================================================================
%==================================================================
%==================================================================

%==================================================================
%==================================================================
 %==================================================================
 %==================================================================
 
In Figure \ref{fig:rainbow}, there are visible discontinuities in the peak temperature, EM-weighted temperature, EM ratios, and thermal energy at the transitions between intervals with XRT data included vs. not (this effect is most visible in the last two orbits). Adding XRT to the DEM delivers a noticeable increase in the value of the EMD peak temperature (and EM-weighted temperature), as well as a decrease in the percent of the emission measure that lies above the peak (a natural consequence of the former). It also leads to an increase in the total thermal energy of the distribution (arbitrary units), though for the cases in this study the difference is within the uncertainty range defined by the iterative DEM-estimation process. This effect occurred when using both the DEMREG and \texttt{xrt\_dem\_iterative2} methods. 

This effect should be further investigated in future studies where both SXR and HXR diagnostics are available to help characterize the thermal distribution of a source. As the range around the peak of the EMD is where the bulk of the distribution is located, there may be significant effects on the energetics of the thermal plasma in a case where the change in input instruments modifies the EMD beyond the uncertainties defined by the DEM process. 

\section{DEMs and the \nustar\ Gain Discrepancy }\label{gainagain}

Spectroscopy of HXR microflares observed by \nustar\ in this and other active regions has revealed consistent discrepancies in observed locations of known solar spectral lines. The analysis in \cite{duncan2021} (Appendix A) concludes that this can be attributed to a small artificial shift which occurs in the \nustar\ gain during observations in the extremely high-count-rate/low-livetime regime ($>10^5$ cps, $<1\%$ livetime) experienced by the instrument during observations of bright solar sources. No gain discrepancy has been observed during observations of standard astrophysical sources, which are several orders of magnitude fainter. 

The solar-specific gain discrepancy causes artificially lower energies to be observed by \nustar\ in comparison to the actual incident spectrum (for example, the strong 6.7 keV Fe complex in solar flares is often observed by \nustar\ at $\sim$6.4 keV). A gain correction procedure is outlined in \cite{duncan2021} for use in HXR spectroscopy. For the events in that study (some of which are the same transients covered in this analysis), this resulted in a multiplicative (gain slope) correction to observed energies on the order of a few percent, which was applied to spectroscopy of the events during their rise/peak intervals. 

For DEMs of most microflares, the process of correcting \nustar\ inputs for a gain discrepancy is straightforward: first, \nustar\ spectroscopy should be performed of the source time interval and region planned for use as a DEM input. A gain correction can be applied to the \nustar\ spectrum via the procedure described in \cite{duncan2021}. This corrected spectrum can then be used to find the \nustar\ emission observed in each energy range of interest (the \nustar\ data input to the DEM). 

\begin{figure*}[!ht] 
\centering 
\includegraphics[width=0.85\textwidth]{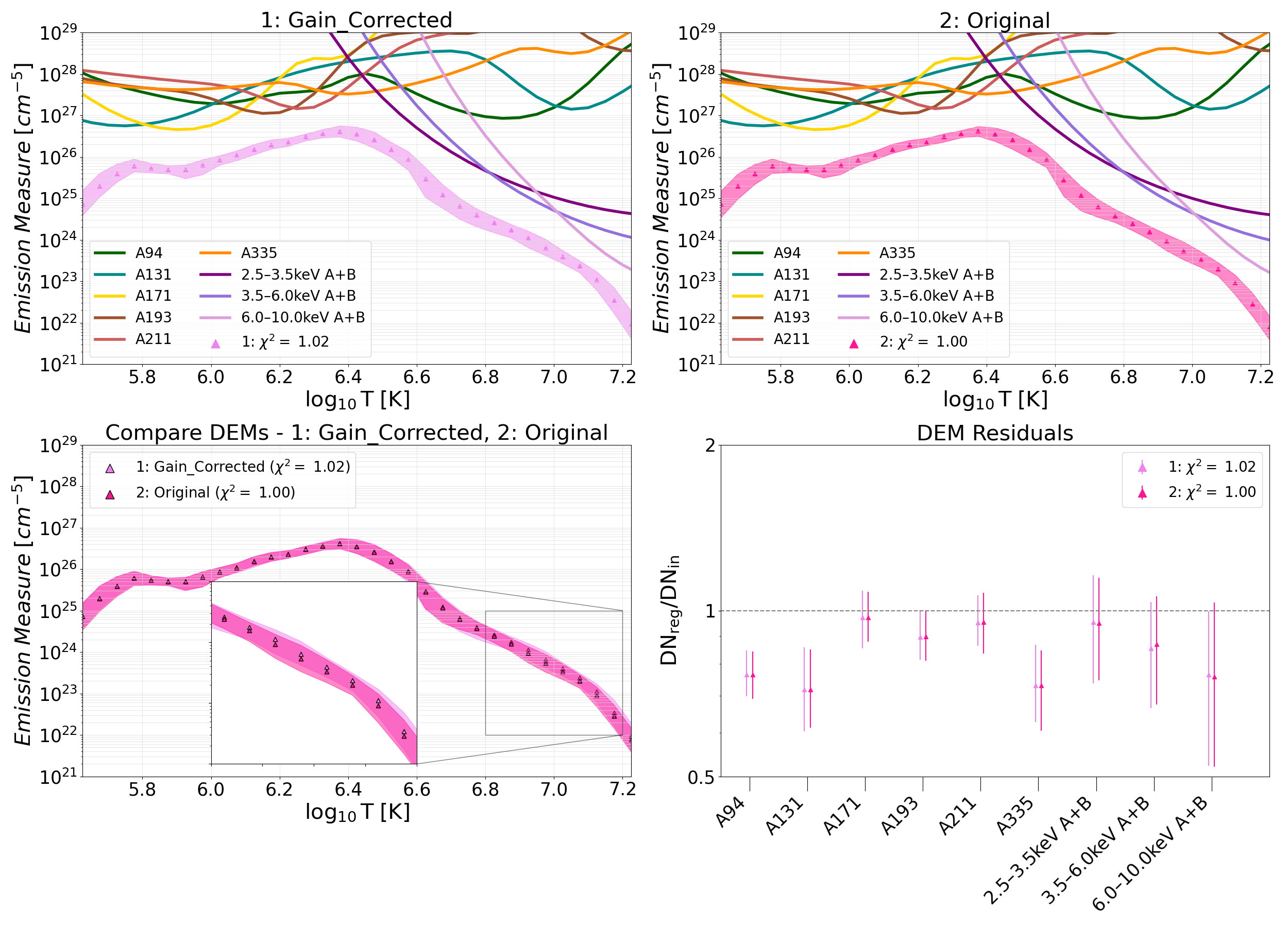} 
\caption{Comparison of DEM results during the rise of a microflare (16:18:15–16:18:45 UTC) with and without applying a correction to account for \nustar's gain discrepancy. The EMDs are consistent with each other within their uncertainties over the full interval, with the largest differences noticeable at higher temperatures (see discussion in Appendix \ref{gainagain}). } 
\label{fig:gainfig}
 \end{figure*}

Figure \ref{fig:gainfig} shows the results from applying a gain correction to an interval during the rise phase of the largest microflare. We see that the few-\% correction in the energies of the observed \nustar\ emission has a minimal effect on the EMD, with corrected and original cases consistent within their mutual uncertainty ranges. In comparison to the original EMD, the corrected distribution predicts 8\% more total emission measure above 5 MK (and 16\% (19\%) more above 7 (10) MK). We emphasize that because the observed gain discrepancy involves a shift to artificially \textit{lower} energies, analysis using uncorrected \nustar\ data when this discrepancy is occurring will always find \textit{less} hot material than is actually present. 

At times when the 6.7 keV Fe complex is not a prominent feature in the \nustar\ spectrum (as is often the case at quiet times), the influence of any possible gain discrepancy is not straightforward to determine or correct. A small gain discrepancy acting on the bremsstrahlung continuum would  closely resemble a similar continuum source with slightly modified temperature and emission measure. 

Because this analysis includes quiet times where the established gain correction procedure is not possible, and because the test gain correction in \ref{fig:gainfig} showed a minimal effect on the EMD, a gain correction is not included in the results presented in this study. However, there may well be cases (likely, DEMs of larger microflares) where a gain correction will be necessary for DEM analysis involving \nustar. 

\bibliography{DEMrefs}{}
\bibliographystyle{aasjournal}

\end{document}